\documentstyle[prb,aps,twocolumn]{revtex}

\begin{document}
\draft
\title{Tunneling Between Two-Dimensional Electron Gases in a
Weak Magnetic Field}
\author{N. Turner, J.~T. Nicholls, K.~M. Brown, E.~H. Linfield,
M. Pepper, D.~A. Ritchie, and G.~A.~C. Jones}
\address{Cavendish Laboratory, Madingley Road, Cambridge CB3 0HE,
England}
\date{\today}
\maketitle

\begin{abstract}
We have measured the tunneling between two two-dimensional
electron gases (2DEGs) in weak magnetic fields,
when the carrier densities of the two electron layers are matched.
At zero magnetic field, $B=0$, the lineshape of the
equilibrium tunneling resonance is best fit by a Lorentzian,
with a linewidth which
is determined by the roughness of the tunnel barrier.
For $B\not=0$ with filling factors $\nu\gg 1$, there is a suppression
of the resonant
tunneling conductance about zero bias,
$V_{\text{sd}}=0$.
This low field signature of the high field Coulomb gap shows the same
linear $B$ dependence as previously measured for $\nu<1$.
\end{abstract}


\narrowtext

Recent
experimental\cite{eis92a,eis94a,brown94c,brown94e}
and
theoretical\cite{yang93,hats93,song93a,johan93,efros93,varma94}
studies of
tunneling between two-dimensional electron gases (2DEGs) have
concentrated on the high magnetic field regime. At fields
when the filling factor $\nu < 1$, there is a strong suppression of
the resonant equilibrium tunneling which is observed at zero field.
This suppression has been labelled
the ``Coulomb gap'' and is thought to be caused
by electron-electron
interactions.\cite{eis92a}
Various theoretical
descriptions\cite{yang93,hats93,song93a,johan93,efros93,varma94,aleiner94}
of the Coulomb gap
are only valid in certain regimes,
conveniently described by relationships between
three lengths:
the magnetic length $l_B$ (determined
by the applied magnetic field $B$, $l_B=\sqrt{\hbar/eB}$),
the average in-plane electron spacing $a$ (fixed by
the carrier density $n$, $a\propto n^{-1/2}$), and the distance
$d$ between the centers of the two quantum wells
(set during  sample growth).

In an earlier paper\cite{brown94c} we explored the
high magnetic field regime,
$\nu < 1$, when $l_B/a$ is (by definition) less than unity,
measuring the gap width as a function of $n$ and $B$.
In experiments at both constant $n$ and fixed $\nu=1/2$,
we observed\cite{brown94c} a high field gap which was independent
of carrier density,
and depended linearly on $B$.
These results do not agree with  many-body theories  which
predict that the energy  gap should scale
with either $e^2/\epsilon l_B \sim \sqrt{B}$
(Refs.~\onlinecite{hats93,song93a,efros93})
or $e^2/\epsilon a \sim \sqrt{n}$
(Refs.~\onlinecite{johan93,varma94}).
Recently Eisenstein {\em et al.\/}\cite{eis95} have also
measured high field $I$--$V$
characteristics at $\nu=1/2$, investigating the gap as a
function of $d$ and $a$.
At high fields electrons can be considered
to be  point-like,
and the interlayer electron-hole pair created by a
tunneling event is expected to be reasonably long-lived.
The width of the gap was argued\cite{eis95}
to be determined by $e^2/\epsilon a$ modified by a smaller
excitonic term proportional to $-e^2/\epsilon d$.
Their experiments show evidence for an interlayer exciton
formed by the tunneled electron and the hole left behind.
It is also proposed\cite{eis95} that the average voltage position
$\langle V\rangle$ of the tunneling current peak should
scale as some function $f$ of $\nu$, as
$\langle V\rangle = f(\nu)\times (e^2/\epsilon a-e^2/\epsilon d)$.

It has been suggested\cite{eis95} that our high field
data,\cite{brown94c} that we have described as having a
linear $B$ dependence, might better be described by a
$\sqrt{B}$ dependence offset from the origin.
In this paper we present data for higher filling factors
$\nu > 1$ which corroborates
our previous measurements\cite{brown94c} for $\nu<1$, and
distinguishes conclusively between gaps with $\sqrt{B}$ and
linear $B$ dependencies.
We show that for layers with matched carrier densities
there is a suppression of tunneling, where the
width of the gap depends {\em linearly on $B$ alone\/}
over the range of filling factors $0.4<\nu <10$.
In our samples we see no evidence for a
change in the functional form of the width of the gap
as we move from $l_B/a<1$ to $l_B/a \gg 1$.

The results presented in this paper are from
measurements on a sample from wafer C751
(the same wafer, but a different sample to that used
in Ref.~\onlinecite{brown94c}).
The two GaAs quantum wells are $180\;\text{\AA}$ wide, and
separated by a $125\;\text{\AA}$ wide
$\text{Al}_{0.33}\text{Ga}_{0.67}\text{As}$ barrier.
The as-grown
carrier densities and mobilities of the upper and lower 2DEGs are
$n_1=3.1 \times 10^{11}\;\text{cm}^{-2}$,
$\mu_1=8 \times 10^5\;\text{cm}^2/\text{Vs}$ and
$n_2=1.8 \times 10^{11}\;\text{cm}^{-2}$,
$\mu_2=2 \times 10^5\;\text{cm}^2/\text{Vs}$.
The carrier densities in the
$100\;\mu\text{m} \times 150\;\mu\text{m}$
tunneling area are controlled by voltages $V_{\text{g}_1}$
and $V_{\text{g}_2}$ to a
front gate $800\;\text{\AA}$ above the upper 2DEG,
and a buried backgate $3500\;\text{\AA}$ below the lower 2DEG.
The proximity of the backgate
to the lower 2DEG,
compared with a backgate evaporated onto a thinned sample,
allows us to accurately control $n_2$ and ensure that
the carrier density is uniform over the entire tunneling area.
References~\onlinecite{lin93,brown94a,brown94b} give details
of the design and fabrication of our devices.

Before discussing data taken in a magnetic field, we
present zero-field tunneling measurements of our sample.
Figure~\ref{zerob} shows the differential
conductance $G=dI/dV_{\text{sd}}$ as a function of top gate voltage
$V_{\text{g}_1}$,
when the carrier density in the lower well was fixed
at $n_2=1.6 \times 10^{11}\;\text{cm}^{-2}$.
Resonant tunneling between
the two 2DEGs occurs when their carrier densities are equal,
$n_1=n_2$,
corresponding to simultaneous conservation
of energy and momentum of the tunneling electrons.\cite{eis91a}
After subtracting a weak parabolic background
$(0.15-4.2V_{\text{g}_1}+10V_{\text{g}_1}^2)$,
the gate sweep characteristic in Fig.~\ref{zerob}
has been fit to the Lorentzian lineshape
\begin{equation}
G(V_{\text{g}_1})=\frac{G_0}{1+\left(\frac{V_{\text{g}_1}-V_0}
{\delta\! V_{\text{g}}}\right)^2},
\label{eqlorentz}
\end{equation}
with $V_0=-0.1968\;\text{V}$,
$\delta\! V_{\text{g}}=0.01874\;\text{V}$,
and $G_0 = 6.70\;\mu\text{S}$.
$V_0$ is  the gate voltage at which $n_1=n_2$; the resonance
can be shifted to a different value of $V_0$
by applying a different backgate voltage $V_{\text{g}_2}$,
thereby changing the carrier density $n_2$ of the lower 2DEG.
We have also tried fitting the resonance to the derivative of a
Fermi function
(${\cal F}\sim\text{sech}^2[e(V_{\text{g}_1}-V_0)/2kT]$)
and a Gaussian.
Of the three functional forms, a Gaussian fit is the poorest,
$\cal F$ is considerably better,
but the overall shape, and especially the tails of the resonance are
best fit by a Lorentzian.
Further evidence that $\cal F$
is inappropriate comes from the
temperature dependence of the resonance width;
the width of $\cal F$ scales with $kT$,
whereas the measured width of the resonance
changes little between $50\;\text{mK}$ and $4.2\;\text{K}$.

The shape of the resonance as a function of $V_{\text{sd}}$,
the DC bias applied between the two layers,
makes us
more confident of our Lorentzian fit.
Assuming that equilibrium tunneling is described by
Eq.~\ref{eqlorentz},
the non-equilibrium conductance $G(V_{\text{sd}})$
with matched carrier densities is calculated to be
\begin{equation}
G(V_{\text{sd}})=G_0 \frac{1-\left(\frac{eV_{\text{sd}}}
{\delta\! E_{\text{g}}}\right)^2}
{\left(1+\left(\frac{eV_{\text{sd}}}
{\delta\! E_{\text{g}}}\right)^2\right)^2}.
\label{noneqbm}
\end{equation}
Figure~\ref{gvsd} shows a comparison of this function with
experimental
data at $B=0$.
The form of $G(V_{\text{sd}})$ calculated from $\cal F$
shows deeper regions of negative differential conductance and
fits the experimental data less well than
Eq.~\ref{noneqbm} derived from a Lorentzian.
The full width at half maximum (FWHM) of Eq.~\ref{noneqbm} is
$1.06 \delta\! E_{\text{g}}$ (which we will call $2 \delta\! E_V$)
and can be compared with the FWHM ($2 \delta\! V_{\text{g}}$)
of Eq.~\ref{eqlorentz},
after the latter has been converted to energy units via the relation
$\delta\! E_{\text{g}} = \delta\! V_{\text{g}}
(d E_{\text{F}} / d V_{\text{g}})$.
{}From Figs.~\ref{zerob} and~\ref{gvsd} we measure
$\delta\! E_{\text{g}}\approx 0.54\;\text{meV}$ and
$\delta\! E_V\approx 0.32\;\text{meV}$.
Eisenstein {\em et al.\/}\cite{eis92a} have measured
$\delta\! E_V \approx 0.3\;\text{meV}$
in a much higher mobility sample, having
$\mu\approx 3\times 10^6\;\text{cm}^2/\text{Vs}$.
The close agreement of our value of $\delta\! E_V$ with
other measurements,\cite{fnote}
even though our
2DEGs are more disordered by a factor of 10--20,
leads us to believe that the linewidth is not caused by
scattering processes within either 2DEG.

The lineshape for equilibrium tunneling has been
calculated\cite{zheng93a}
to be Lorentzian, under the assumption that the
width is determined by the ``quantum lifetime'' of the electrons,
$\delta\! E=\hbar/\tau$, where $\tau$ is a lifetime
derived from disorder scattering.
The experimental width $\delta\! E_{\text{g}}$ of our resonance
corresponds to a quantum lifetime of $\tau\approx 1.2\;\text{ps}$.
The transport lifetime in
the more disordered 2DEG is
$\tau_{\text{tr}} = m^\ast\mu/e \approx 7.6\;\text{ps}$.
Experimentally we find the width of the resonance peak
(and hence $\tau$) to be independent of $n$ over
the range $0.6<n<3.5\times 10^{11}\;\text{cm}^{-2}$,
a fact which is at odds with
theoretical predictions\cite{gold88}
for the $n$ dependence of both the transport and single-particle
relaxation (small-angle scattering) times in 2D.
If scattering is predominantly due to remote
ionised impurities, the scattering times are predicted to vary as
$\tau_{\text{tr}}\sim n^{3/2}$ and $\tau_{\text{sp}}\sim n^{1/2}$.
As the linewidth does not depend on the in-plane properties of
the 2DEGs, the most likely broadening mechanism is
non-uniformity in the width of the tunnel barrier.
If the well width $w$ of $180\;\text{\AA}$ varies by up to
one monolayer ($\delta\! w\approx\pm 3\;\text{\AA}$)
the broadening $\delta\! E_0$ of the
ground state energy $E_0$, assuming a hard-wall square-well
potential, will be
$\delta\! E_0 = E_0 \times 2\delta\! w/w \approx 0.44\;\text{meV}$,
which is consistent with the measured value
of $\delta\! E_{\text{g}}\approx 0.54\;\text{meV}$.

Resonant tunneling between 2DEGs in a magnetic field is
expected to occur when Landau levels (LLs) of the same index are
aligned in the two wells.
For wells with matched carrier densities, the resonance
should occur at
$V_{\text{sd}}=0$.
Figure~\ref{gvsd} shows typical $G(V_{\text{sd}})$ data,
when the carrier densities in the two layers were matched at
$n_{1,2}=0.95 \times 10^{11}\;\text{cm}^{-2}$, at
four magnetic fields; for clarity the traces are offset vertically.
The $B=0$ data is plotted as points and compared with
Eq.~\ref{noneqbm},
shown as a solid line.
There is a single peak centered about $V_{\text{sd}}=0$, with
a width determined by the scattering time $\hbar / \delta\! E_V$,
surrounded by
regions of weak negative differential conductance.
At $B=0.9\;\text{T}$ (corresponding to $\nu=4.3$) there is a clear
suppression of tunneling at $V_{\text{sd}}=0$, the single
zero-field peak has split into two distinct peaks.
The top trace, taken at $B=2.6\;\text{T}$, $\nu=1.5$,
shows an increased
separation of the split peaks and $G(V_{\text{sd}}=0)$ is reduced.
The suppression of the zero bias conductance is a manifestation
of the Coulomb gap, eventually causing
$G(V_{\text{sd}}=0)$ to vanish when $\nu<1$.

The splitting $\Delta$ of the conductance peaks shown in
Fig.~\ref{gvsd} is a measure of the width of the Coulomb gap.
Moreover, $\Delta$ defined in this way is
equivalent to the gap parameter
in the expression $I=I_0 \exp(-\Delta/V_{\text{sd}})$,
introduced by He {\em et al.\/}\cite{song93a} and which
has been used\cite{eis94a,brown94c} to fit
high field $I$--$V_{\text{sd}}$
characteristics.
Figure~\ref{deltab} shows $\Delta$ measured from
peak-to-peak splittings as a function of $B$,
showing a basically linear
behaviour over most of the range of magnetic field,
except at particular values of magnetic field.

The measured values of $\Delta$ at $B=0.9$ and $2.6\;\text{T}$
are in good agreement with the linear fit of Fig.~\ref{deltab},
but the value of $\Delta$ from the
curve in Fig.~\ref{gvsd} at $B=0.4\;\text{T}$
($\nu=9.8$) is too large.
As the LL filling approaches  integral and certain
fractional (2/3, 1/3) values
the simple picture of tunneling from one
bulk 2DEG to another is distorted by the
presence of edge state transport in the two layers,
and our understanding of the gap is complicated.
As these filling factors are approached
the magnitude of the
tunneling conductance is
reduced and the measured value of $\Delta$ increases
rapidly with $B$, producing the bumps identified in
Fig.~\ref{deltab}.
The dashed line in Fig.~\ref{deltab} is a least-squares straight
line fit $\Delta(\text{meV})= 0.45\hbar\omega_{\text{c}}-0.19$
which ignores points (open circles) affected by the
presence of edge states.
The high field data in  Fig.~\ref{deltab} reproduces our
previous\cite{brown94c,brown94e} work for $\nu<1$,
and in combination with
the low field data in this paper convincingly supports
a linear $B$ dependence over a wide magnetic field range,
in spite of the complications
caused by the apparent hardening of the gap
at integral filling factors.


Previously, using only the high-field data, we were unable to
determine whether the linear fit passed through the
origin, or through some point close to the origin
(the error in the intercept parameter being
very much larger than the value of the parameter itself).
It is clear from Fig.~\ref{deltab} that the high field
points have considerably less scatter than those at
low fields.
This makes any link between the negative intercept and a
screened exciton binding energy slightly tentative.

In the only available theory\cite{aleiner94}
predicting a low field ($l_B/a\gg 1$) gap,
it is assumed that the
tunneling electron acquires extra Coulomb energy
$e^2 / \epsilon a$ on
entering the electron liquid.
However, this is not the final state of the system
and the increased local charge density relaxes back to its background
value by the emission of virtual magnetoplasmons.
Lorentz forces acting on the spreading
carriers cause the
formation of a stable vortex of energy
\begin{equation}
E_0=\frac{\hbar\omega_{\text{c}}}{2\nu}\ln
\left(\nu\frac{e^2}{\epsilon\hbar v_{\text{F}}}\right),
\label{aleinereq}
\end{equation}
where
$v_{\text{F}}$ is the Fermi velocity.
The energy $E_0$ sets the scale for the gap,
and Eq.~\ref{aleinereq}
(ignoring the weak logarithm) has a quadratic dependence
on magnetic field at fixed $n$, $E_0 \sim B^2$.
In contrast, our experimental results show that the low $B$ gap
is linear in magnetic field, with the same constant of
proportionality as that measured at high fields.
In a slightly different system
Ashoori {\em et al.\/}\cite{ash90,ash93} have measured
the temperature dependence of the tunneling
conductance deduced from the capacitance.
They observe
a suppression of tunneling which depends only on $B$,
independent of LL filling,
and  deduce the gap
width from the activation of the conductance.
Their results were interpreted using a model with a tunneling
density of states with a linear
gap at the Fermi energy,
with a magnitude of
$0.047\hbar\omega_{\text{c}}$, one tenth of our measured value.
The measurements presented in this paper
differ from these previous experiments\cite{ash90,ash93}
in the low $B$ regime,
in that the tunneling reported here is 2D--2D tunneling
(from one 2DEG to
another) as opposed to 2D--3D (from a 2DEG to an
$n^+$ bulk layer).
Therefore, it might be expected
that we should measure a wider
gap (if there is a gap in both the emitter and collector
density of states) than Ashoori {\em et al.\/}

In principle inter-LL tunneling, between LLs in the
two 2DEGs with different indices, is
forbidden.
In practice scattering of electrons within the tunnel
barrier makes these transitions weakly allowed.
For wells with matched carrier densities, this gives rise to
conductance peaks at
$eV_{\text{sd}}=\pm\hbar\omega_{\text{c}},\:
\pm2\hbar\omega_{\text{c}}\ldots$
where $\omega_{\text{c}}=eB/m^\ast$ is the cyclotron frequency.
In addition to the high field suppression of tunneling,
it has been observed\cite{eis92a,brown94c}
that the inter-LL tunneling peaks were shifted to slightly
higher energies than the usual $\hbar\omega_{\text{c}}$ spacing.
The downward pointing arrows in Fig.~\ref{gvsd} mark the
positions of the inter-LL tunneling peaks.
For fields in the range $0.5<B<3\;\text{T}$
the separation of the inter-LL peaks in the two bias directions
is measured to be $2.2 \hbar \omega_{\text{c}}$,
which is larger than
the expected value of $2\hbar\omega_{\text{c}}$.
At high fields, the inter-LL tunneling has been
observed\cite{eis92a,brown94c} as a peak in $I$--$V_{\text{sd}}$
characteristics at
$eV_{\text{sd}}\approx 1.3 \hbar\omega_{\text{c}}$.
The equivalent measurement\cite{brown94c,brown94e}
in conductance $dI/dV_{\text{sd}}$ has
a peak at $eV_{\text{sd}} \approx 1.1 \hbar\omega_{\text{c}}$.
Here we observe an inter-LL tunneling peak at the
enhanced value of $eV_{\text{sd}}=1.1\hbar\omega_{\text{c}}$
all the way down to $B=0.5\;\text{T}$, below which inter-LL
tunneling is not clearly discernible.
In the interpretation that the enhanced LL spacing
is due to a reduced
effective mass, our observations would suggest
a constant mass reduction
which is independent of filling factor, in contradiction with the
predictions of
Smith {\em et al.\/}\cite{smith92}

In conclusion, we have shown that the electron
tunneling lifetime is governed by
the interface roughness of the tunnel barrier,
rather than by scattering within either of the 2DEGs.
While we have seen some weak evidence that
the linear gap does not pass through the origin
($\Delta(B=0)\not=0$),
and hence that there may be some excitonic effects,
we have presented much stronger evidence
that the linear relationship $\Delta \sim B$
(rather than $\sqrt{B}$ or $B^2$) fits the data
over a wide field regime,
supporting our previous measurements\cite{brown94c,brown94e}
and contradicting available theories.

We wish to thank the Engineering and Physical Sciences
Research Council (UK) for supporting this work.
JTN acknowledges support from the Isaac Newton Trust,
and DAR acknowledges support from Toshiba Cambridge Research Centre.


\begin{figure}
\caption{Differential conductance $G(V_{\text{g}_1})$,
where the gate voltage $V_{\text{g}_1}$ controls the
top 2DEG carrier density $n_1$. The carrier density
of the bottom 2DEG was fixed
at $n_2=1.6 \times 10^{11}\;\text{cm}^{-2}$.
After subtracting a weak parabolic background from
the raw data,
the tunneling peak fits the Lorentzian
lineshape given in Eq.~\protect\ref{eqlorentz}.}
\label{zerob}
\end{figure}

\begin{figure}
\caption{Differential tunneling conductance $G(V_{\text{sd}})$
at matched
carrier densities $n_1 = n_2 = 0.95\times 10^{11}\;\text{cm}^{-2}$,
measured at $B=0,\:0.4,\:0.9,\:2.6\;\text{T}$ (the four curves are
offset vertically).
The data for $B=0$ (some points are not shown for clarity)
is plotted with Eq.~\protect\ref{noneqbm} (solid line).
The splitting of the zero field resonance into two
peaks defines the gap width $\Delta$.}
\label{gvsd}
\end{figure}

\begin{figure}
\caption{The measured gap width $\Delta$ plotted versus $B$.
Open circles ($\circ$) show the gap hardening due to the
presence of edge states, solid circles ($\bullet$)
are unaffected by edge states.
The dashed line
is a least-squares fit through the points
unaffected by the presence of edge states,
$\Delta=0.45\hbar\omega_{\text{c}}-0.19$.}
\label{deltab}
\end{figure}

%

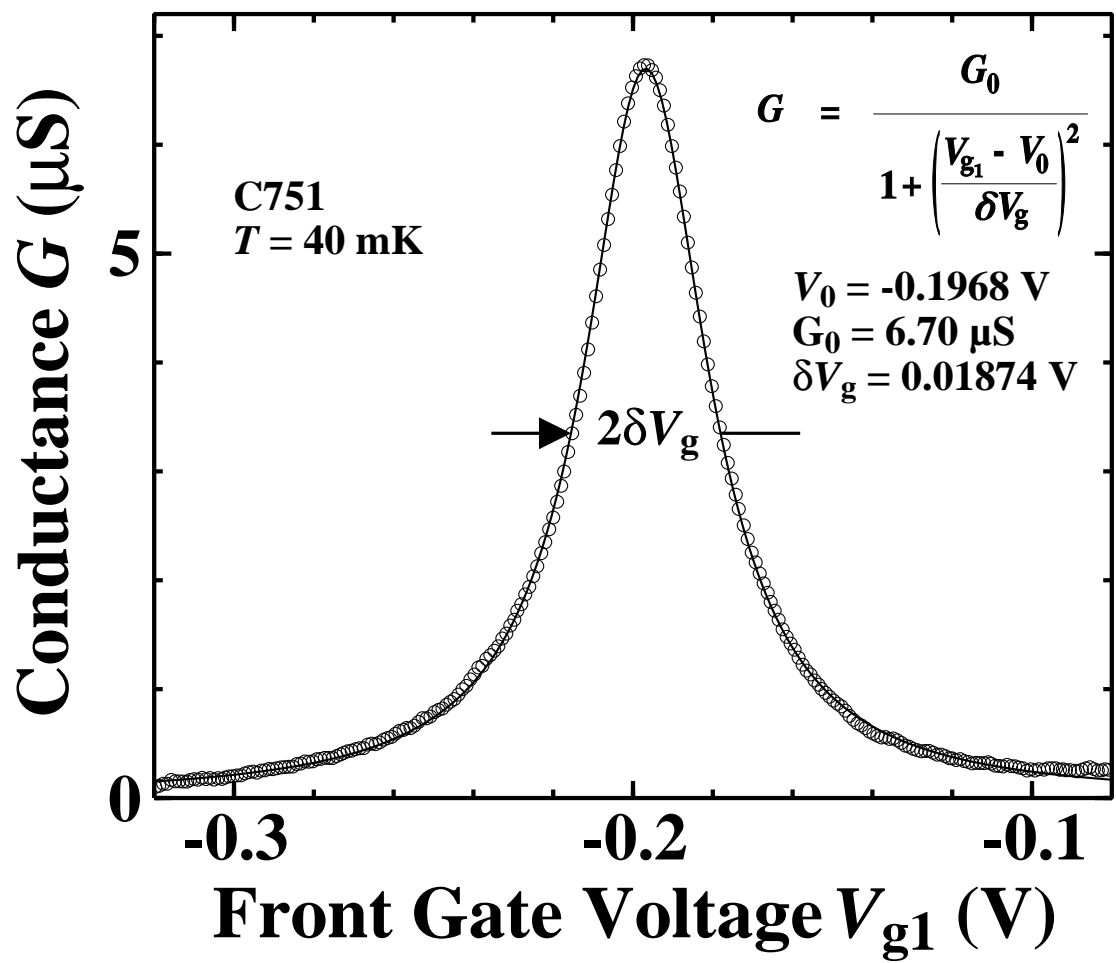
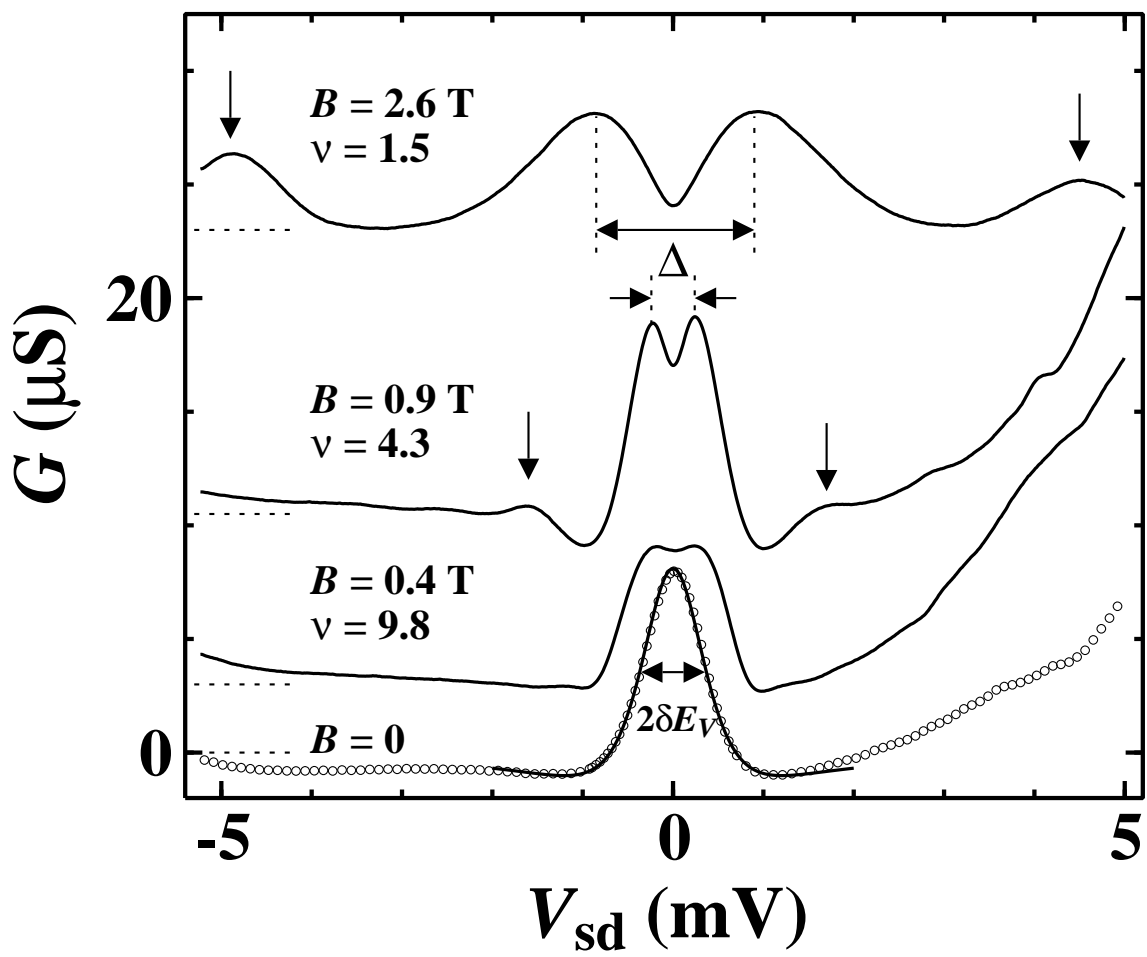
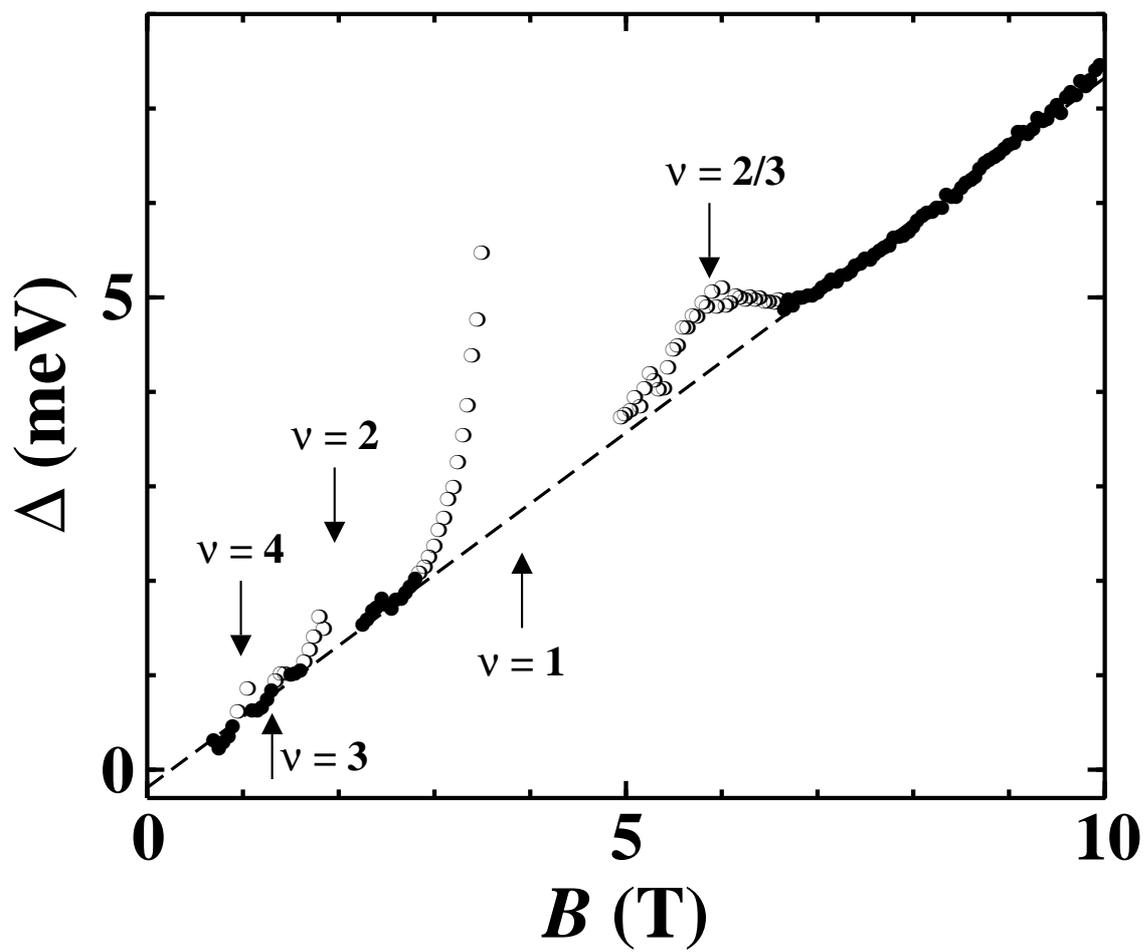
\end{document}